\documentclass[a4paper,11pt]{article}
\usepackage{pos}
\usepackage{mathrsfs}
\newcommand{\keuso}{\mbox{K-EUSO}}

\title{%
Expected performance of the \keuso{} space-based observatory
}
 \ShortTitle{%
 Expected performance of the \keuso{} space-based observatory
 }

\author*[a]{Francesco Fenu}
\author[b]{Sergei Sharakin}
\author[b]{Mikhail Zotov}
\author[c]{Naoto Sakaki}
\author[c]{Yoshiyuki Takizawa}
\author[a]{Marta Bianciotto}
\author[a]{Mario Bertaina}
\author[c]{Marco Casolino}
\author[b]{Pavel Klimov}

\affiliation[a]{Università degli studi di Torino,\\
  Via Pietro Giuria 1 10125, Torino, Italy}

\affiliation[b]{Skobeltsyn Institute of Nuclear Physics,\\
  Lomonosov Moscow State University, Russia}

\affiliation[c]{RIKEN,\\
Wako, Saitama, Japan}

\affiliation[d]{INFN, Structure of Rome Tor Vergata\\
Rome, Italy}

\forColl{JEM-EUSO}

\emailAdd{francesco.fenu@gmail.com}

\abstract{\keuso{} is a planned mission of the JEM-EUSO program for the
study of ultra-high energy cosmic rays (UHECR) from space, to be
deployed on the International Space Station.  The \keuso{} observatory
consists of a UV telescope with a wide field of view, which aims at the
detection of fluorescence light emitted by extensive air showers (EAS)
in the atmosphere.  The EAS events will be sampled with a time
resolution of 1--2.5~$\mu$s to reconstruct the entire shower profile
with high precision.  The detector consisting of $\sim 10^{5}$
independent pixels will allow a spatial resolution of $\sim$700~m on
ground. From a 400~km altitude, \keuso{} will achieve a large  and full sky exposure to sample the highest energy range of the UHECR spectrum. In this
contribution, we present estimates of the performance of the
observatory: an estimation of the expected exposure and triggered event
rate as a function of energy and the event reconstruction performance,
including resolution of arrival directions and energy of UHECRs.

}

\FullConference{37$^{\rm{th}}$ International Cosmic Ray Conference (ICRC 2021)\\
		July 12th -- 23rd, 2021\\
		Online -- Berlin, Germany}

\begin{document}
\maketitle

\section{Introduction}

The nature and origin of ultra-high energy cosmic rays (UHECRs) remains unsolved in contemporary astrophysics.
The very low fluxes at extreme energies, of the order of few particles per km$^2$~sr per century above $\sim$50 EeV, are a challenge for current observatories.
The main goal of the Joint Experiment Missions
for Extreme Universe Space Observatory (\mbox{JEM-EUSO}) program is the
investigation of the UHECRs of the most extreme energies through the
detection of fluorescence and Cherenkov light emitted by extensive air
showers in the atmosphere~\cite{JEMEUSO_Program}.  From several hundred
 kilometers of altitude, such wide field of view
telescopes will allow a very large exposure and therefore will probe the
most extreme part of the spectrum.

\begin{figure}[!ht]
	\centering \includegraphics[height=5cm]{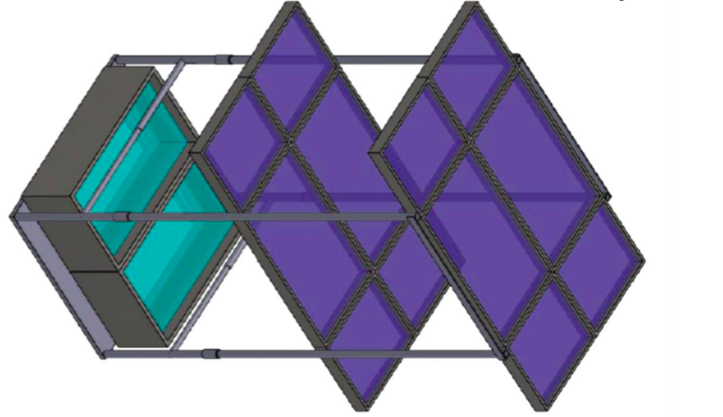}
	\caption{Scheme of the simulated \keuso{} detector.}
	\label{fig:scheme}
\end{figure}

\keuso{} is a fundamental cornerstone of the JEM-EUSO program.
It is the first mission in this framework which will be capable of UHECR
detection from space.  It is planned to fly in 2024 and to be placed onboard the
Russian segment of the International Space Station.  The design presented here is a modified version of what was shown in \cite{CasolinoPTEP} and is developed to fit the size and weight constraints imposed by the location on the International Space Station. The detector
consists of a refractive optical system of $1400\times2400$~mm$^{2}$ size (see
Fig.~\ref{fig:scheme}).  The optics consists of two Fresnel lenses that
focus the light onto a focal surface of $1300\times1000$~mm$^{2}$ size.  The
focal surface consists of 44 Photo Detector Modules (PDMs), of 36 Hamamatsu R11265-103-M64 Multi Anode Photomultipliers
(MAPMTs) each.
A MAPMT consists of 64 independent channels (8 per side) with a
3~mm size.  Each of these channels (identified as pixels in the
following) has a field of view of 0.1$^\circ$ which corresponds to
$\sim700$~m on the ground.  The time resolution is in the process of
definition and ranges from 1~$\mu$s to 2.5~$\mu$s.  
This parameter will be determined as a trade-off between the
limited hardware and telemetry budgets and the need of a good time
resolution.

The quantum efficiency of the MAPMTs is between
35 to 40\% in the wavelength range 300--400~nm.  The photomultiplier
signal is read out and amplified by the SPACIROC3 ASIC~\cite{spaciroc3}
already in use in Mini-EUSO~\cite{Mini-EUSO-general}.  The SPACIROC3
operates in the single photoelectron mode and has a double pulse resolution
of less than 10~ns.  The majority of the data handling tasks, like data
buffering, configuration of the read-out ASICS, triggering,
synchronization and interfacing with the CPU system, is performed by the
Zynq board.  Given the very high time resolution of the detector
($\sim1$~MHz) and the high number of pixels, no full data retrieval is
possible.  Data must therefore satisfy strict trigger conditions.
Concentrations of the signal are sought for by trigger algorithms to
preferentially select the shower signal while rejecting background
events.
Data are saved on a hard drive, to be flown back to Earth, with a subset
being  sent by telemetry, as is being done now for
Mini-EUSO.

\begin{figure}[!ht]
\centering
\includegraphics[height=6.5cm]{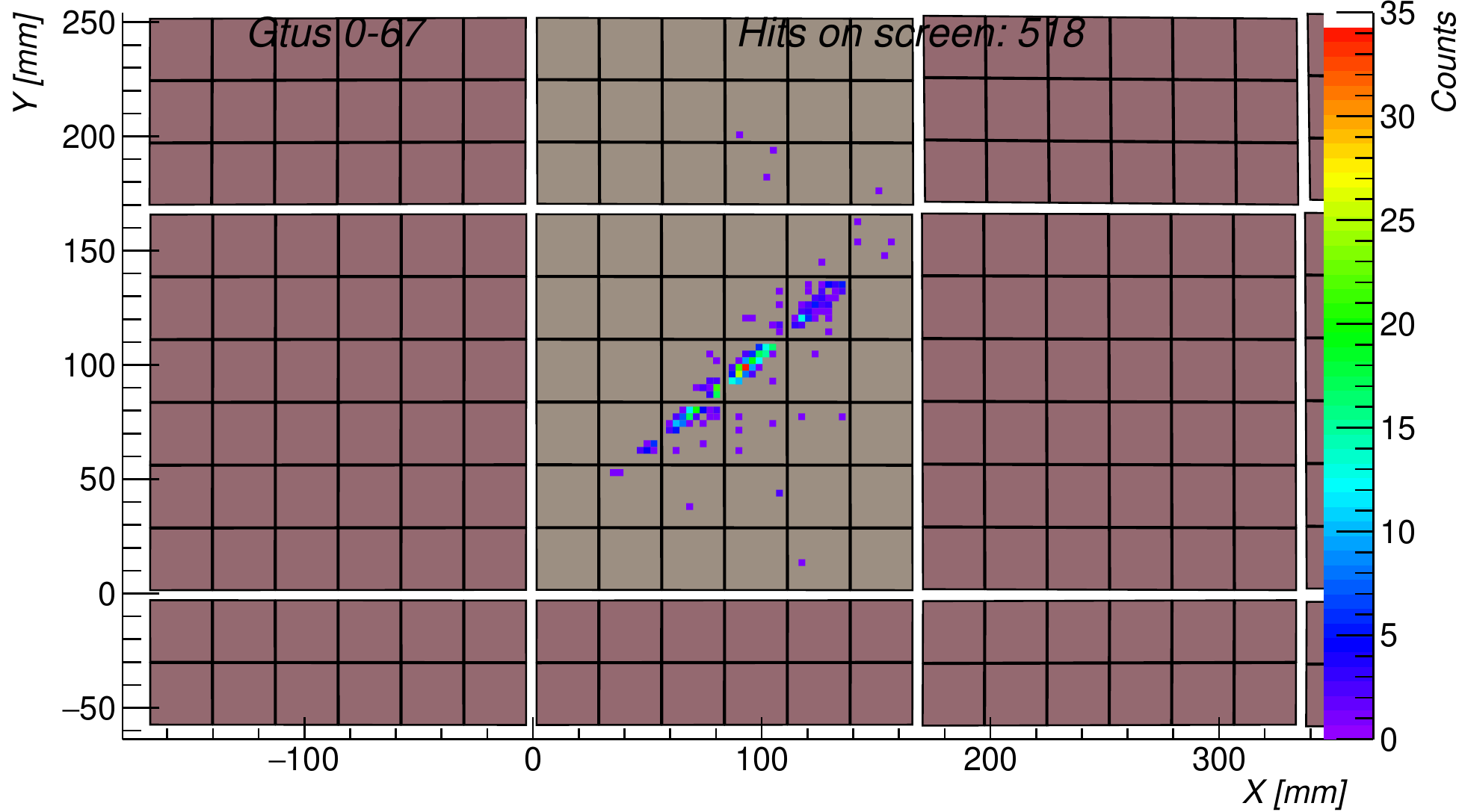}

\includegraphics[width=.48\textwidth]{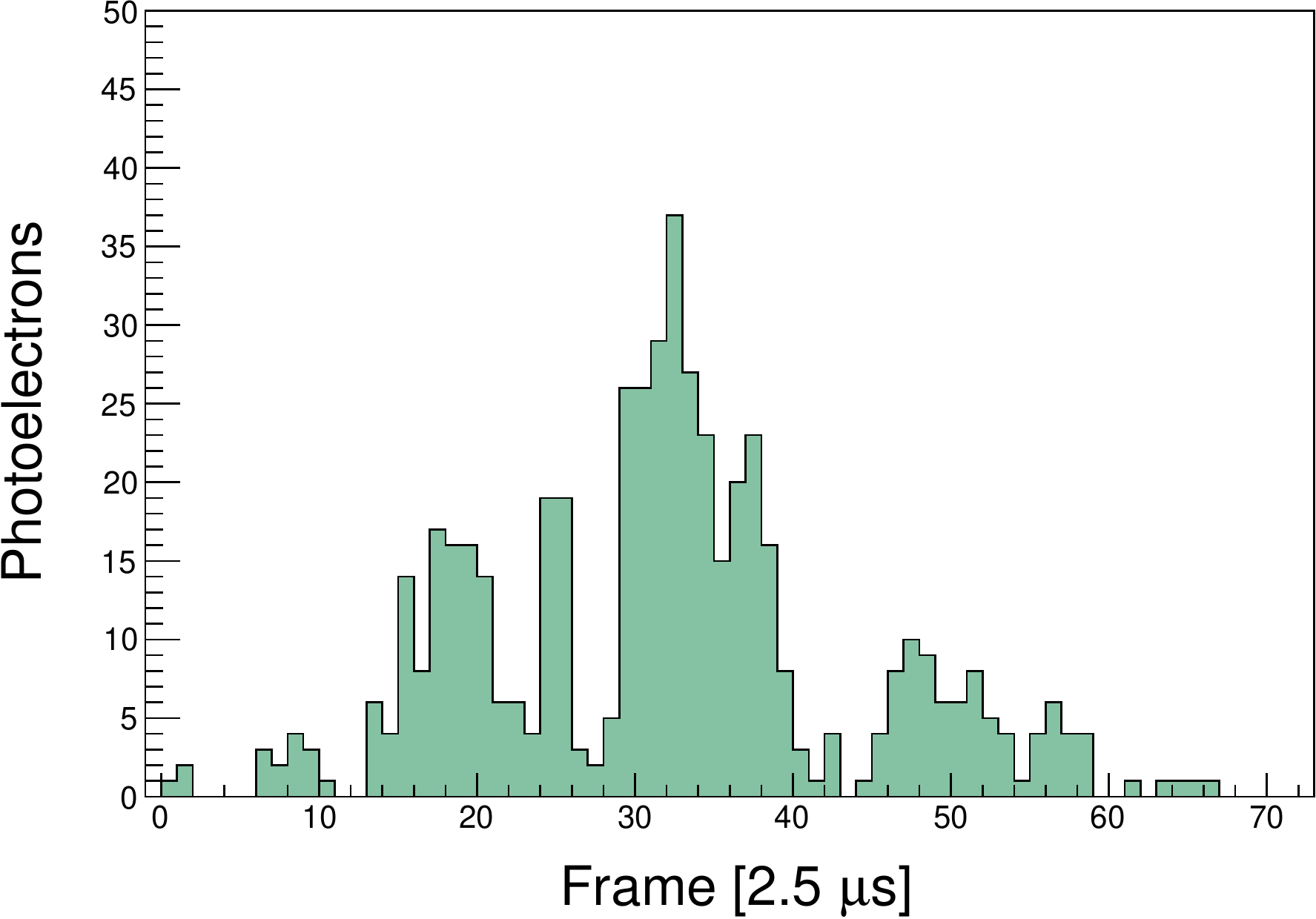}\quad
\includegraphics[width=.48\textwidth]{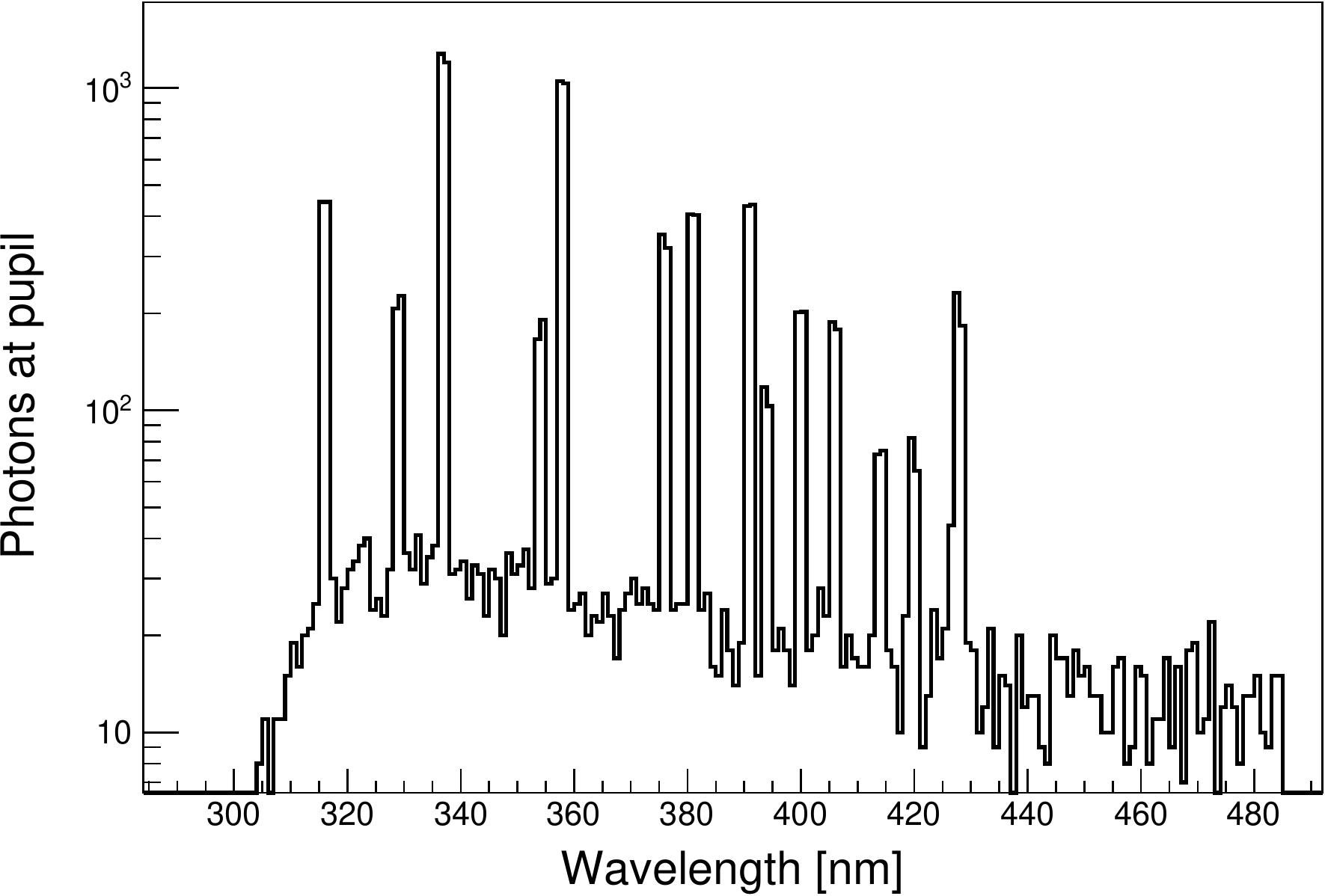}

\caption{A $10^{20}$ eV, 60$^\circ$ event simulated by ESAF. Only signal
	from the shower is shown.  On the top: the distribution of the signal
	generated by the shower on the focal surface.  On the left: the
	corresponding simulated photoelectron time distribution. On the
	right: the spectrum of photons at the detector entrance.}

\label{image:ESAF1e20_60}
\end{figure}

In this work, we estimate the scientific performance of the \keuso{}
detector. The study is done by means of simulations performed with the
ESAF framework~\cite{berat}.  An example of a cosmic ray event simulated
by ESAF is shown in Fig.~\ref{image:ESAF1e20_60}. The top panel presents
an example of a simulated photoelectron distribution of a cosmic ray
shower (without any airglow emission).  On the bottom left panel the
same photoelectrons are plotted as a function of time.  It can be seen
that empty areas between MAPMTs on the focal surface cause a periodic
decrease in the signal intensity.  The bottom right panel shows the
wavelength spectrum of the photons entering the detector.  The
fluorescence emission lines can be seen together with the Cherenkov
continuum emission. 

\section{The \keuso{} exposure}

The trigger algorithms of the \keuso{} mission have been developed in
the framework of the JEM-EUSO program \cite{trigger} and are currently in the process of
optimization.  The logic is structured in multiple stages, each reducing
the trigger rate by several orders of magnitude.  The first level
trigger, to be operated at the level of the PDM, looks
for concentrations of the signal localized in space and time.  The
second level trigger is activated each time the first level trigger
conditions are satisfied and integrates the signal intensity in a
sequence of test directions.  Directions consistent with one of the
simulated extensive air showers have a higher chance of overcoming a
preset threshold. Whenever such condition is met, the second level
trigger is issued.  The activation of the second level trigger starts
the transmission and data storage procedure.  The data acquisition is
therefore stopped and data are either saved on a hard disk or sent to
ground by telemetry.  The trigger therefore reduces the data flow by
several orders of magnitude.  Thresholds are set to have a rate of the
order of a trigger every few seconds at most to make the data
acquisition consistent with the telemetry budget.  The aim of this
section is therefore to test the efficiency curve of the algorithm with
respect to cosmic ray showers.

The exposure calculation is based on a Monte Carlo simulation of EAS of variable energy and direction.  To avoid border effects, cosmic rays are
injected in an area~$A_\mathrm{simu}$ larger than the field of view.
The ratio of the triggered $N_\mathrm{trigg}$ over simulated events
$N_\mathrm{simu}$ is then calculated for each energy bin. The solid
angle~$\Omega$ from which the cosmic rays arrive on the field of view is
also included in the formula.  The effects of day-night cycle and moon
phases are taken into account in~$\eta$, the astronomical duty cycle.
Effects of clouds and artificial lights are also taken into account
by $\eta_\mathrm{clouds}$ and $\eta_\mathrm{city}$ respectively.
In this formula, we assumed $\eta$ = 0.2, $\eta_\mathrm{clouds} = 0.72$
and $\eta_\mathrm{city}$ = 0.9 as estimated in~\cite{Astrop_exposure}. The
exposure~$\mathscr{E} (E)$  is then calculated over time~$t$, which
is assumed to be~1 year in the following:

\begin{equation} \label{eq:exposure}
  \mathscr{E} (E) = \frac{N_\mathrm{trigg}}{N_\mathrm{simu}} (E)
		\times A_\mathrm{simu} \times \Omega \times \eta \times
		\eta_\mathrm{clouds} \times \eta_\mathrm{city} \times t.
\end{equation}

\begin{figure}[!ht]
\centering
\includegraphics[height=6.5cm]{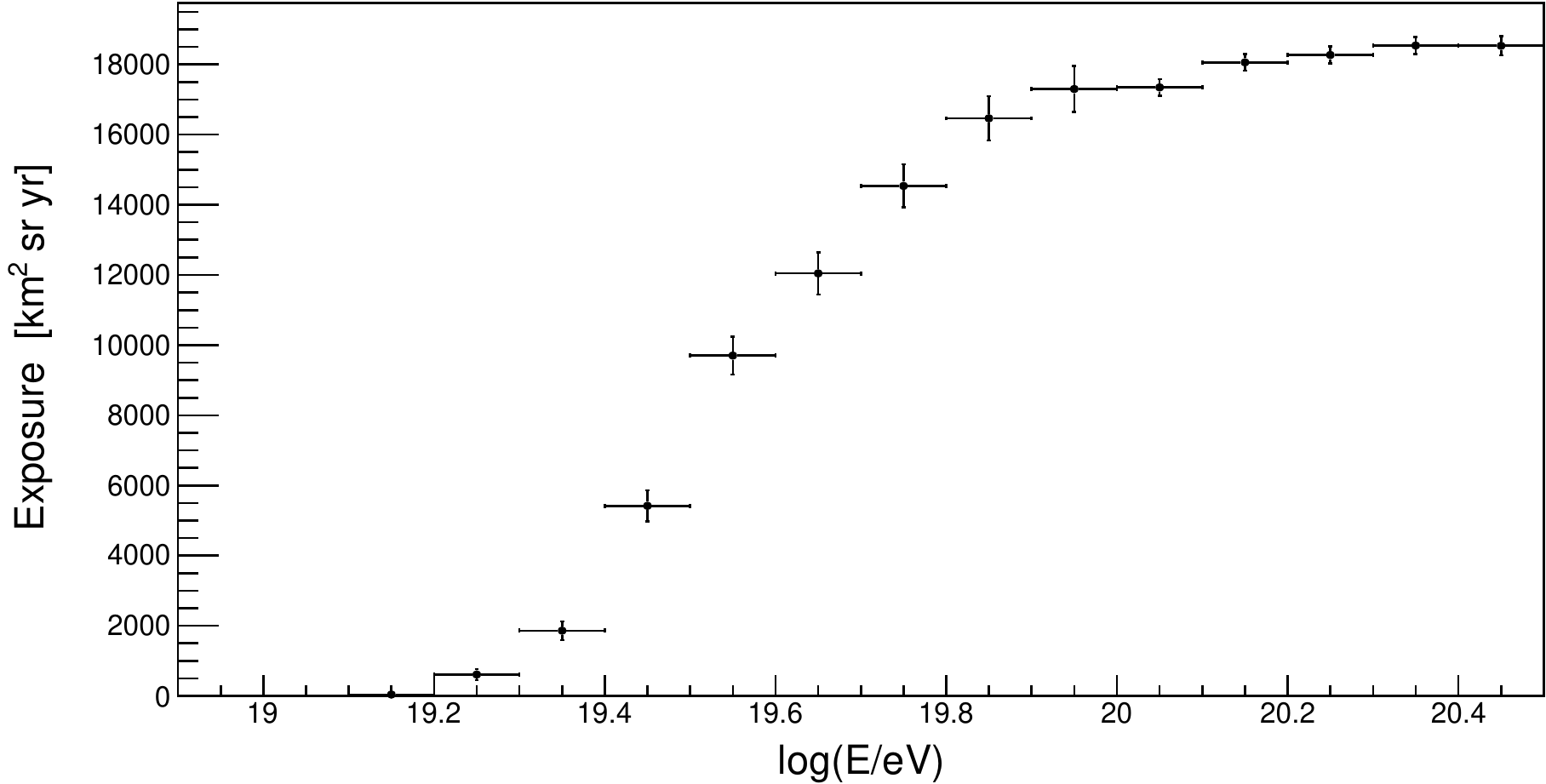}

\caption{Annual exposure of \keuso{} as a function of UHECR primary energy.}
\label{fig:exposure}
\end{figure}

The yearly exposure as a function of energy is shown in
Fig.~\ref{fig:exposure}.  As it can be seen, at the plateau, which is
reached at around $10^{20}$~eV, \keuso{} achieves an exposure of
$\sim18000$~km$^2$~sr~yr.  The 50\% efficiency is reached around $\sim 4
\times 10^{19}$~eV.

Assuming the spectrum recently published by the Auger
collaboration~\cite{AugerPRD}, the expected rate of triggered events has
been calculated to be of the order of 4 events/year above $10^{20}$~eV
and 65 events/year above $5 \times 10^{19}$~eV.

\section{The \keuso{} angular reconstruction}

At the occurrence of a trigger, the acquisition is stopped and data are
retrieved.  The information collected at this point is used then to
reconstruct the parameters of the primary particle.

The first step consists in the recognition of the track, namely of
pixels and frames in which the light of the shower arrives.  A
comprehensive review of the signal identification methods is given
in~\cite{EA_Angle} and~\cite{EA_Energy}.  All the methods look for
concentrations of the signal in space and time that display kinematics
consistent with the one of an extensive air shower.

\begin{figure}[!ht]
	\centering
	\includegraphics[width=.48\textwidth]{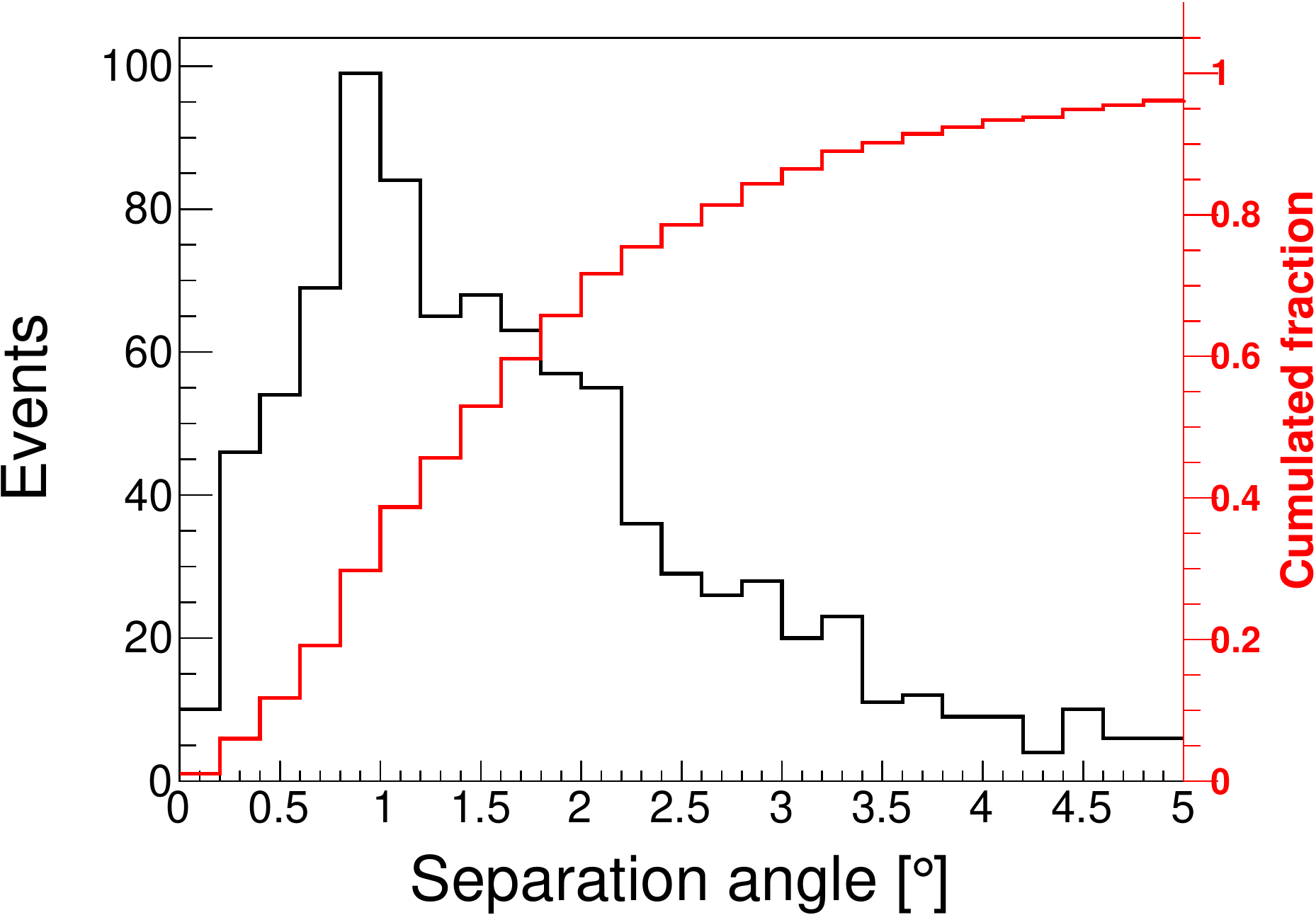}\quad
	\includegraphics[width=.48\textwidth]{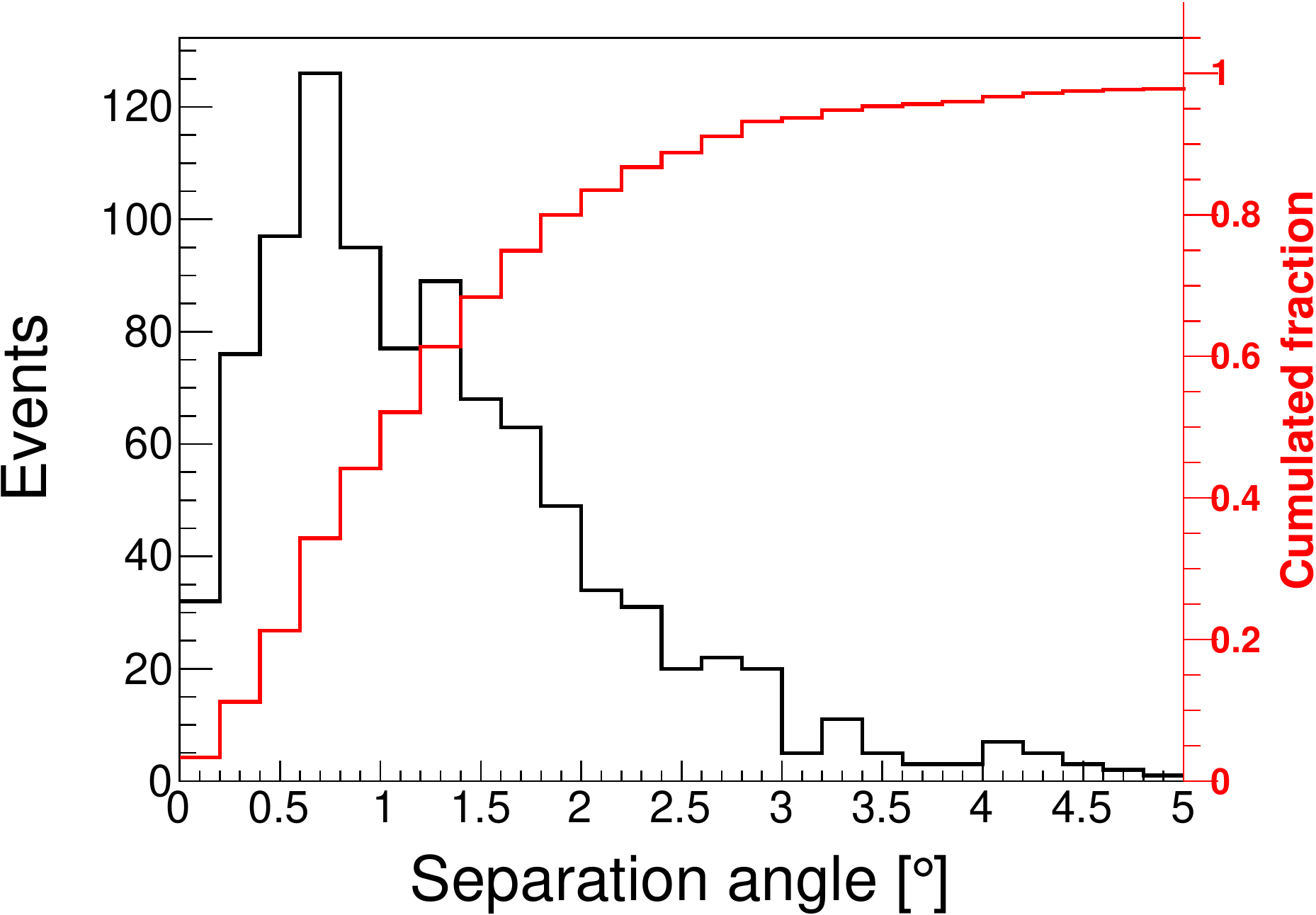}

	\caption{Performance of the angular reconstruction for UHECRs
	with the energy $10^{20}$~eV in
	the center of the field of view. Left: 45$^\circ$ zenith angle.
	Right: 60$^\circ$ zenith angle.}

	\label{fig:AngleReco}

\end{figure}

The angular reconstruction extracts the arrival direction of the primary
particle from the distribution and timing of the identified track.
Several methods have been tested in the context of the JEM-EUSO
program~\cite{EA_Angle}.  The method used for this work is based on a
$\chi^2$ fit of the position and timing of the shower signal (the so
called \textit{Numerical Exact~1} method).  In this method the
identification of the plane where both the shower and the detector lie is the
first step of the procedure.  The zenith angle of the shower is then
reconstructed by comparing the arrival time of the photons from a test
shower and the identified track.  In Fig.~\ref{fig:AngleReco}, we show
the reconstruction performance for arrival directions of
EAS of $10^{20}$~eV and zenith
angles equal to ~$45^\circ$
and $60^\circ$ in the center of the field of view.  To assess the
quality of the reconstruction we plot the integral of the event
distribution from~0 to a specific angle (in red).

\begin{figure}[!ht]
	\centering
	\includegraphics[height=7cm]{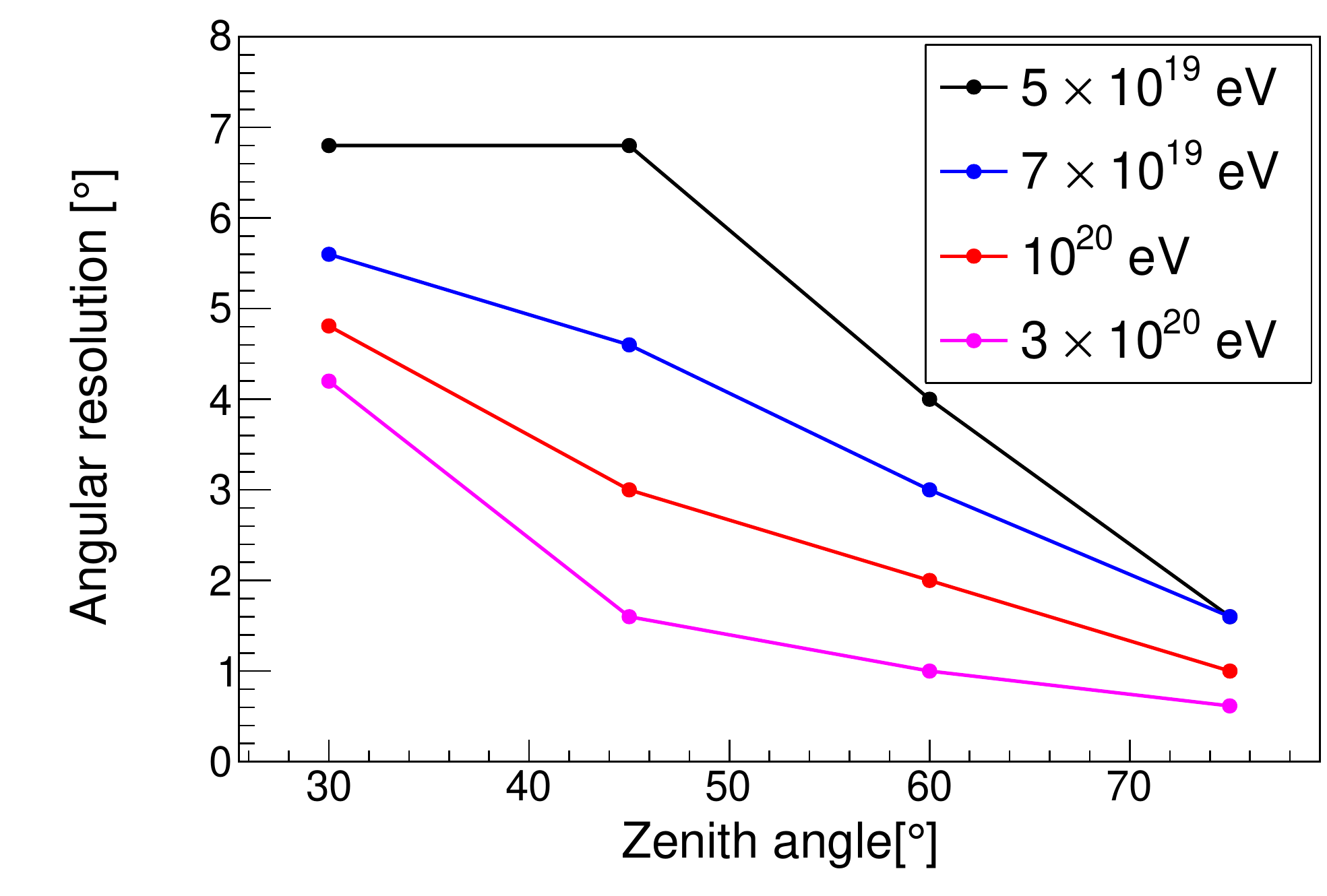}

	\caption{Estimated angular resolution of \keuso{} for different
	zenith angles and energies of a primary particle
	on the full field of view of the detector.}

	\label{fig:AngleRes_ALL_FULL}
\end{figure}

In Fig.~\ref{fig:AngleRes_ALL_FULL} the angular resolution is
plotted as the angle within which 68\% of the events fall.
For this plot, we simulated 500 EASs in 16 different combinations of energy and
zenith angle.  For each condition, the events have been simulated over the
entire field of view of the detector.  It can be seen that \keuso{}
achieves a resolution between 3 to 7 degrees at small zenith angles and
improves to 1--2 degrees for nearly-horizontal events.  There is a clear
improvement trend as the energy increases.

\section{The \keuso{} energy reconstruction}

\begin{figure}[!ht]

	\centering
	\includegraphics[height=7cm]{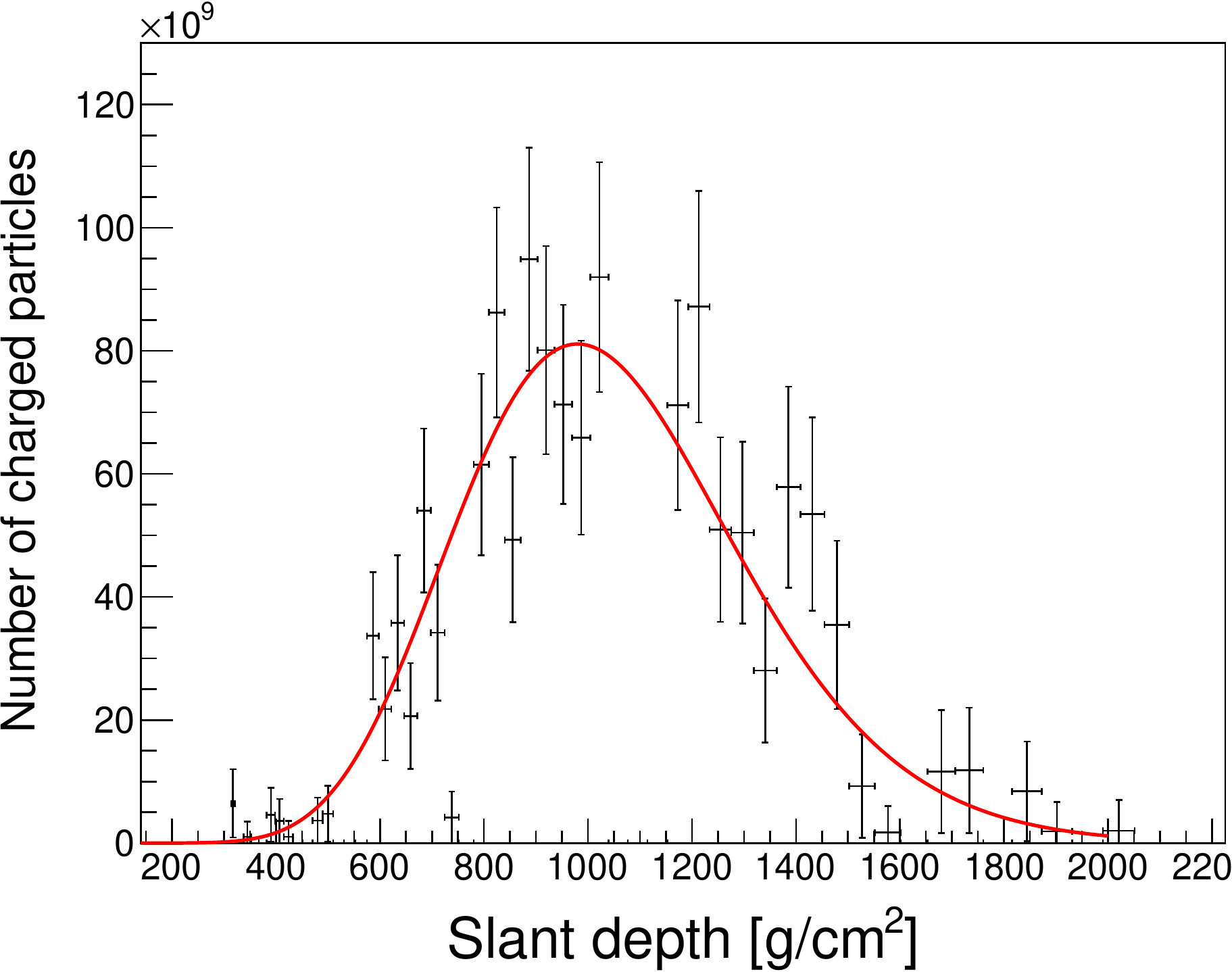}

	\caption{Reconstructed shower profile as a function of slant depth
	is shown as black crosses.  The shower profile fit is
	shown in red.  A shower generated by a $10^{20}$~eV, 60 degrees UHECR
	was simulated.  The reconstructed energy equals $1.22 \times 10^{20}$~eV.}

	\label{fig:Profile}

\end{figure}

The energy reconstruction is done according to \cite{EA_Energy} and is
based on the signal identified by pattern recognition.  The
photoelectron profile is reconstructed based on the counts falling in
the identified track with the airglow subtracted on average.  The
attenuation occurring in the detector is corrected following a look-up
table relating an incident direction and wavelength of photons to the
detector efficiency.  Several methods to reconstruct the shower geometry
have been implemented and are discussed in~\cite{EA_Angle}.  With an
estimate of the position of the shower in atmosphere, it is possible to
calculate the amount of atmospheric extinction and the luminosity curve.
An estimate of the fluorescence yield is then used to reconstruct the
charged particle profile of an EAS.  Such profile is then fit with a
shower profile parameterization to obtain the energy and the depth of
maximum.
An example of a $10^{20}$~eV, 60 degrees
event reconstructed profile is shown in Fig.~\ref{fig:Profile} as black
crosses.  A fit of the shower parameterization is shown in red.
The reconstructed energy was $1.22 \times 10^{20}$~eV.

\begin{figure}[ht!]
	\centering
	\includegraphics[width=.48\textwidth]{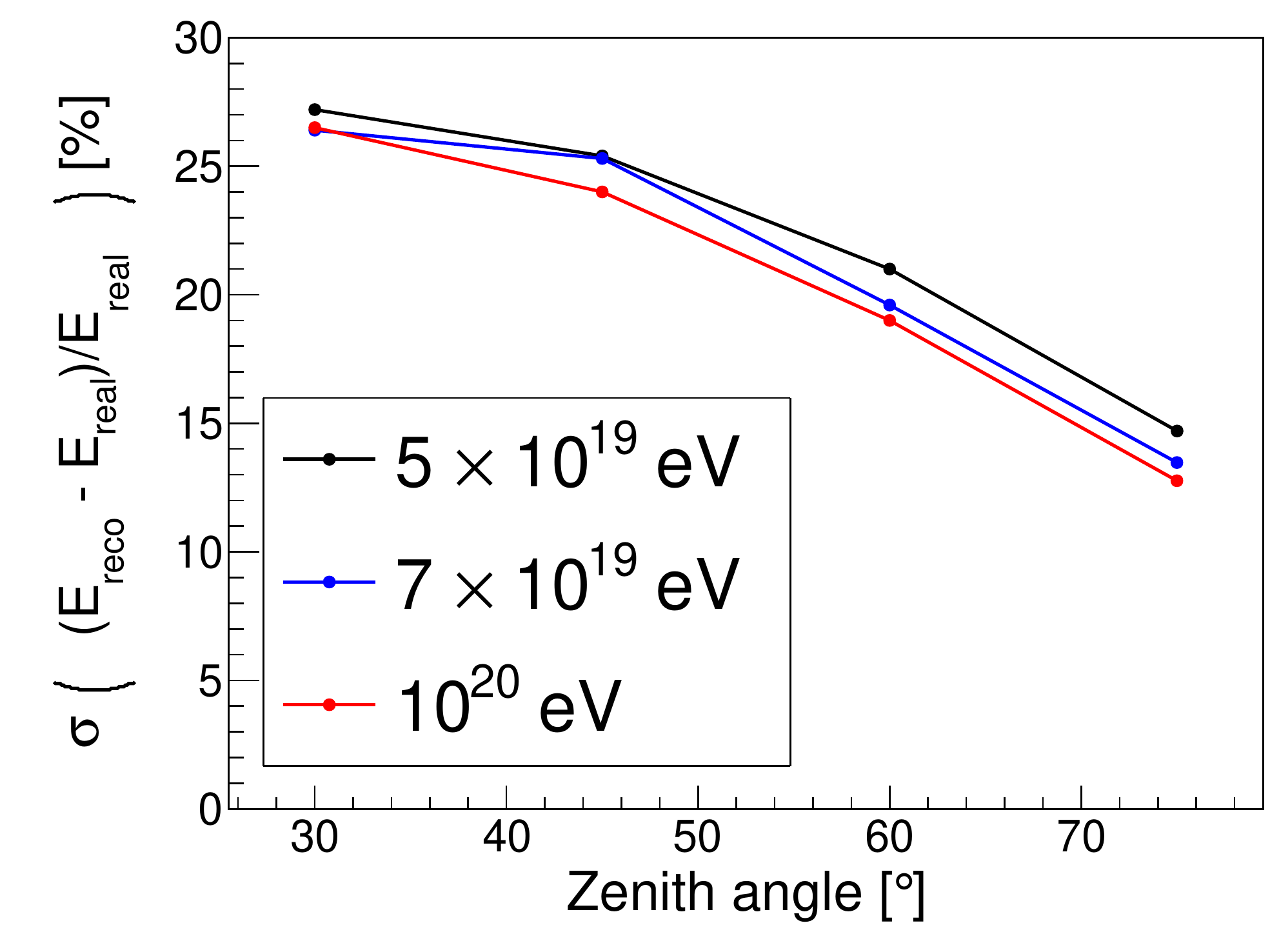}\quad
	\includegraphics[width=.48\textwidth]{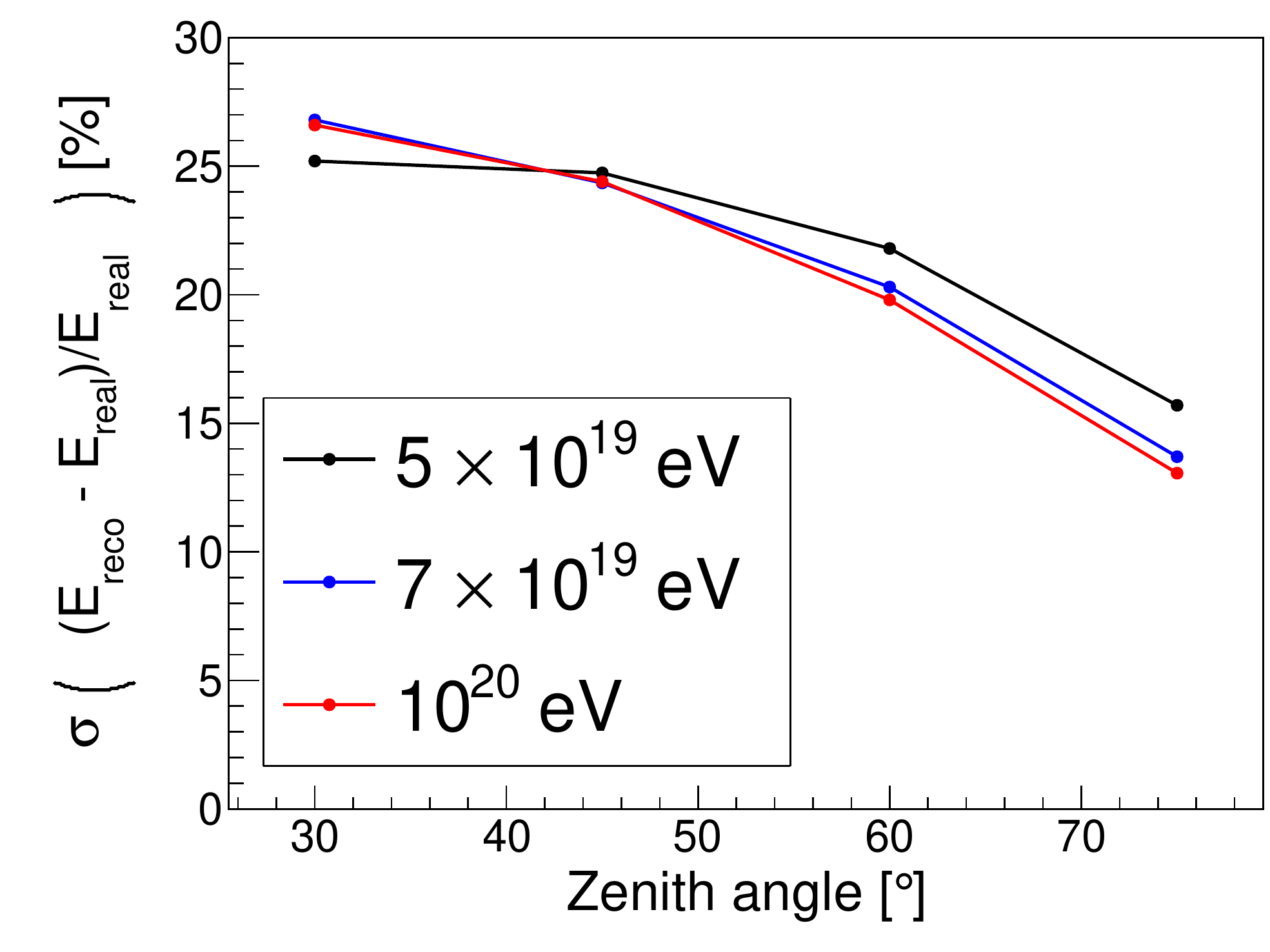}

	\caption{Estimates of energy reconstruction for different energy and
	zenith angle. Left: center of the field of view. Right: full field of
	view. }

	\label{fig:EnergyRes}
\end{figure}

Estimations of the energy resolution of \keuso{} for UHECRs with
different energies arriving at various zenith angles are shown in
Fig.~\ref{fig:EnergyRes}. As in the previous section, 2500 showers were
simulated at fixed energies and zenith angles, both for the center and for
the full field of view of the detector.  The resolution was estimated as
the standard deviation of the  $(E_\mathrm{reco} -E_\mathrm{real}) /
E_\mathrm{real}$ distribution.  It can be seen that the energy
resolution is around 25$\%$ at low zenith angles and improves to
around 15\% for nearly horizontal events, with a small improvement trend toward
higher energies\footnote{With the only exception of the $5 \times 10^{19}$ eV, 30 degrees condition. Here, at the threshold, the reconstruction is particularly challenging.}.  No significant decrease of the performances has been
observed if events are simulated on the full field of view.

\section{Conclusions}

We presented estimations of the performance of the \keuso{} mission,
which is a modified version of the ``KLYPVE'' project, aimed to meet
weight and dimension constraints implied by its planned deployment at
the Russian segment of the ISS.  The expected rate of triggered events
is around 65 per year for energies above $5 \times 10^{19}$~eV and
$\sim4$ above $10^{20}$~eV assuming the Pierre Auger Observatory
spectrum.  The angular reconstruction performance has
also been studied for different conditions of energy and zenith angle on
the whole field of view.  The angular reconstruction varies from 3 to 7
degrees at low zenith angles and improves up to 1--2 degrees for nearly
horizontal EAS.  The higher energies are characterized by a better
performance.  The energy reconstruction has been tested for different
conditions both in the center and on the whole field of view.  The
energy resolution was found to be around 25\% at low zenith angles
and 15\% at higher angles.  A mild trend of an improvement of the
reconstruction performance is observed as the  energy increases.

\acknowledgments

The authors express their deep and collegial thanks to the
entire JEM-EUSO program and all its individual members.
The article was prepared based on research materials carried out in
the space experiment ``KLYPVE,'' included in the Long-term program of
Experiments on board the Russian Segment of the ISS.
This work was supported by the State Space Corporation
ROSCOSMOS by MAECI, the Italian Ministry of Foreign Affairs and
International Cooperation, Projects of Major Relevance, by
JSPS KAKENHI Grant (JP19H01915).

\clearpage
\section*{Full Authors List: \Coll\ Collaboration}

\begin{sloppypar}
{\small \noindent
G.~Abdellaoui$^{ah}$,
S.~Abe$^{fq}$,
J.H.~Adams Jr.$^{pd}$,
D.~Allard$^{cb}$,
G.~Alonso$^{md}$,
L.~Anchordoqui$^{pe}$,
A.~Anzalone$^{eh,ed}$,
E.~Arnone$^{ek,el}$,
K.~Asano$^{fe}$,
R.~Attallah$^{ac}$
% ...
}
\end{sloppypar}
\vspace*{.3cm}

{ \footnotesize
\noindent
% ...
$^{ac}$ LPR at Dept. of Physics, Faculty of Sciences, University Badji Mokhtar, Annaba, Algeria \\
% ...
$^{ah}$ Telecom Lab., Faculty of Technology, University Abou Bekr Belkaid, Tlemcen, Algeria \\
% ...
$^{cb}$ Universit\'e de Paris, CNRS, AstroParticule et Cosmologie, F-75013 Paris, France\\
% ...
$^{ed}$ Istituto Nazionale di Fisica Nucleare - Sezione di Catania, Italy\\
% ...
$^{eh}$ INAF - Istituto di Astrofisica Spaziale e Fisica Cosmica di Palermo, Italy\\
% ...
$^{ek}$ Istituto Nazionale di Fisica Nucleare - Sezione di Torino, Italy\\
$^{el}$ Dipartimento di Fisica, Universita' di Torino, Italy\\
% ...
$^{fe}$ Institute for Cosmic Ray Research, University of Tokyo, Kashiwa, Japan\\
% ...
$^{fq}$ Nihon University Chiyoda, Tokyo, Japan\\
% ...
$^{md}$ Universidad Polit\'ecnia de madrid (UPM), Madrid, Spain\\
% ...
$^{pd}$ University of Alabama in Huntsville, Huntsville, AL; USA\\
$^{pe}$ Lehman College, City University of New York (CUNY), NY, USA\\
% ...
}

\end{document}